\documentclass[%
superscriptaddress,
 amsmath,amssymb,
 aps,
 prstab, twocolumn,
]{revtex4-1}

\usepackage{amsmath}
\usepackage{amssymb}
\usepackage{dcolumn}
\usepackage{graphicx}
\usepackage{longtable}
\usepackage{bm}
\usepackage{color}
\usepackage{epsfig}
\usepackage{hyperref}

\begin{document}

\title{Large anomalous Hall conductivity in Weyl ferrimagnet Cs$_{2}$Co$_{3}$S$_4$ predicted by density-functional calculations}

\author{Gang Bahadur Acharya}
\affiliation{Central Department of Physics, Tribhuvan University, Kirtipur, 44613, Kathmandu, Nepal}
\affiliation{Leibniz-IFW Dresden, Helmholtzstr. 20, 01069 Dresden, Germany}
\author{Manuel Richter}
\affiliation{Leibniz-IFW Dresden, Helmholtzstr. 20, 01069 Dresden, Germany}
\affiliation{Dresden Center for Computational Materials Science (DCMS), TU Dresden, D-01062 Dresden, Germany}
\author{Klaus Koepernik}
\affiliation{Leibniz-IFW Dresden, Helmholtzstr. 20, 01069 Dresden, Germany}
\author{Madhav Prasad Ghimire}
\email{madhav.ghimire@cdp.tu.edu.np}
\affiliation{Central Department of Physics, Tribhuvan University, Kirtipur, 44613, Kathmandu, Nepal}
\affiliation{Leibniz-IFW Dresden, Helmholtzstr. 20, 01069 Dresden, Germany}


\begin{abstract}

The identification of topological Weyl semimetals has recently 
gained considerable attention.
Here, we report the results of density-functional theory calculations
regarding the magnetic properties, the electronic structure,
and the intrinsic anomalous Hall conductivity of the title compound,
which was synthesized already 50 years ago but received little attention, hitherto.
We found Cs$_{2}$Co$_{3}$S$_4$ to be a ferrimagnetic half-metal with a
total spin magnetic moment of about 3 $\mu_B$ per formula unit.
It shows energy band gap of 0.36 eV in the majority-spin channel
and a pseudo-gap at the Fermi level in the minority-spin channel.
We identified several sets of low-energy Weyl points and traced their dependence on the direction of magnetization.
The intrinsic anomalous Hall conductivity is predicted to reach a magnitude up to 
500 $\Omega^{-1}$cm$^{-1}$, which is comparable to values obtained
in other celebrated Weyl semimetals.

\end{abstract}

\maketitle

\section{INTRODUCTION}

Dirac semimetals, Weyl semimetals (WSMs), nodal line semimetals, and triple point semimetals are the four main types of topological semimetals. This classification is based on the node distribution in the crystal momentum $k$ space and the related band 
degeneracy.~\cite{zou2019study,moore2010birth,PhysRevLett.98.106803,RevModPhys.82.3045,tokura2019magnetic,hasan2011three,gao2019topological,weng2016topological,kang2020dirac,ye2019haas,RevModPhys.90.015001,PhysRevLett.120.177704,hirayama2018topological,RevModPhys.93.025002,RevModPhys.88.021004} 
If two non-degenerate bands cross or touch in a single, separated point in $k$-space, this is called a Weyl point (WP). 
WPs are topologically protected and are robust against small 
perturbations~\cite{zheng2019tunable} and they exist in pairs with opposing chiral charges.~\cite{NIELSEN198120,NIELSEN1981173,HOSUR2013857}
Non-degenerate bands, which are required for the existence of WPs,
are present if inversion symmetry (IS) or time reversal symmetry (TRS) or both are broken. 

Most of the recent research on WSMs is focused on non-magnetic compounds, where IS is broken such as in the TaAs 
family~\cite{PhysRevB.89.081106,PhysRevX.5.011029,PhysRevX.5.031013, PhysRevLett.116.066802},
in WTe$_2$~\cite{soluyanov2015type,li2017evidence}, in MoTe$_2$~\cite{PhysRevB.92.161107,deng2016experimental,PhysRevX.6.031021},in Mo$_x$W$_{1-x}$Te$_2$~\cite{chang2016prediction,belopolski2016discovery,PhysRevB.94.085127,PhysRevB.105.125138}, 
in TaIrTe$_4$~\cite{PhysRevB.95.241108,PhysRevB.97.241102}, etc. 
These materials show intrinsic electronic characteristics with potential for future electronic applications. 
For instance, the gapless band structure can be useful for broadband  photodetectors~\cite{wang2017quantum,RevModPhys.90.015001}, 
the spin momentum locking of Fermi arcs for spintronics~\cite{PhysRevB.95.235436}, 
and the helical nature of surface electrons for topological qubits.~\cite{Baireuther_2017,PhysRevLett.120.177704,PhysRevLett.121.237701} 

Studies on magnetic WSMs are still less frequent in comparison to those on non-magnetic WSMs. A pyrochlore iridate, Y$_2$Ir$_2$O$_7$, was the first theoretically predicted 
magnetic WSM~\cite{PhysRevB.83.205101}, followed by HgCr$_2$Se$_4$, 
Ti$_2$MnAl, Co$_3$Sn$_2$S$_2$, YbMnBi$_2$ Co$_2$MnGa, Mn$_3$Sn/Mn$_3$Ge, 
GdPtBi~\cite{PhysRevLett.107.186806,PhysRevB.97.060406,morali2019fermi,
belopolski2019discovery,PhysRevResearch.1.032044,
borisenko2019time,kuroda2017evidence,nayak2016large, hirschberger2016chiral},
as well as certain Co-based Heusler compounds predicted to host a
TRS-breaking Weyl phase.~\cite{Kbler2016,PhysRevLett.117.236401,Kushwaha2018}

In recent years, magnetic WSMs are receiving more attention due to several advantages. For instance, the position of Weyl points in the reciprocal space and their energy can be tuned by rotating the magnetization.~\cite{PhysRevResearch.1.032044} 
In this way, the related intrinsic anomalous Hall conductivity and the anomalous Nernst conductivity are also tuned. 
Further, in WSMs where both IS and TRS are broken, application of an external magnetic field can rotate the magnetization into a general direction such that the Shubnikov group contains only the
identity element. In such a case, pairs of Weyl points split in energy, giving access to chiral electromagnetic responses.~\cite{ray2022tunable} 

The present investigation is devoted to the magnetic compound Cs$_{2}$Co$_{3}$S$_4$.
This compound was first synthesized and characterized by x-ray powder diffraction in 1972.~\cite{bronger1972thiomanganate}
Somewhat later, single crystals of the title compound were prepared by Bronger {\em et al.}~\cite{bronger1988structure}, henceforth referred to as BHM.
The structure and magnetic state of these single crystals were characterized by x-ray diffraction (XRD) and neutron diffraction (ND) in the same work. Though the compound is known for half of a century, there are no further investigations on Cs$_{2}$Co$_{3}$S$_4$ published, to the best of our knowledge. This lack of information motivated us to perform a systematic research on the magnetic ground state and electronic properties of Cs$_{2}$Co$_{3}$S$_4$ on the basis of density functional theory (DFT). We found a ferrimagnetic ground state which is (almost) half metallic and, at the same time, semi-metallic. As a consequence, we could characterize the compound as a WSM and evaluated the intrinsic contribution to the anomalous Hall conductivity (AHC).

\section{CRYSTAL STRUCTURES AND COMPUTATIONAL DETAILS}

\begin{figure*}[ht]
\psfig{figure=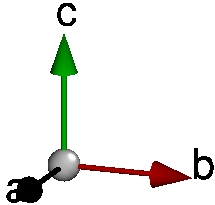,scale=0.28}
\psfig{figure=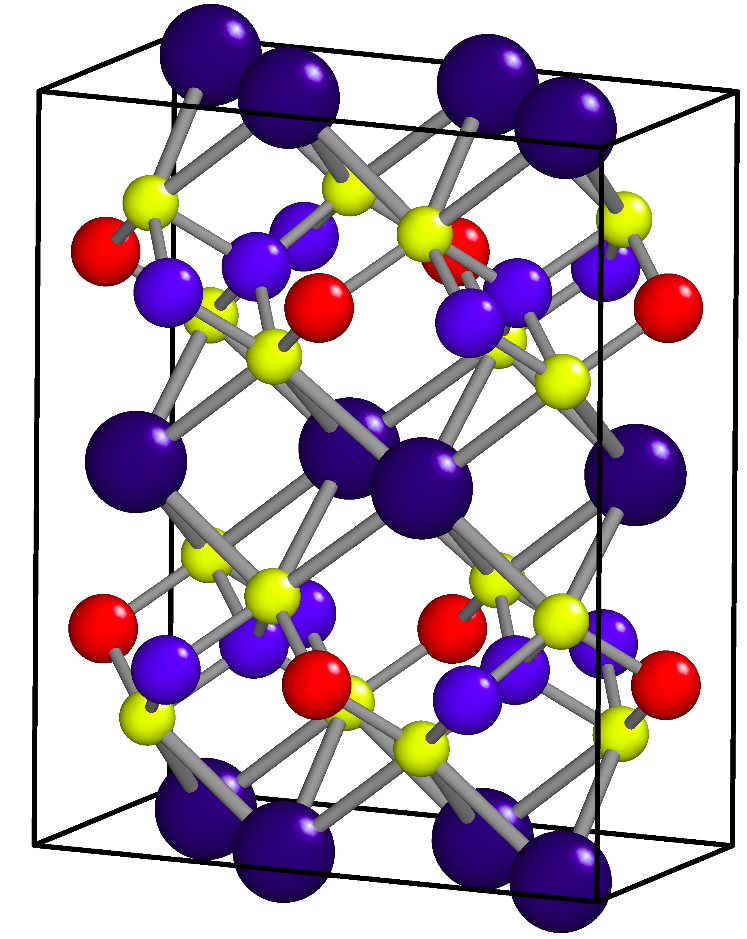,scale=0.16}
\caption{Crystal structure of Cs$_2$Co$_3$S$_4$ in the conventional elementary cell. Violet (blue, red, yellow) balls denote Cs (Co(I), Co(II), S) atoms.}
  	\label{fig:crystal}
  \end{figure*}

\begin{figure*}
  \centering
  \psfig{figure=Fig1a.png,scale=0.22}
  \hspace{-0.1cm}
  \psfig{figure=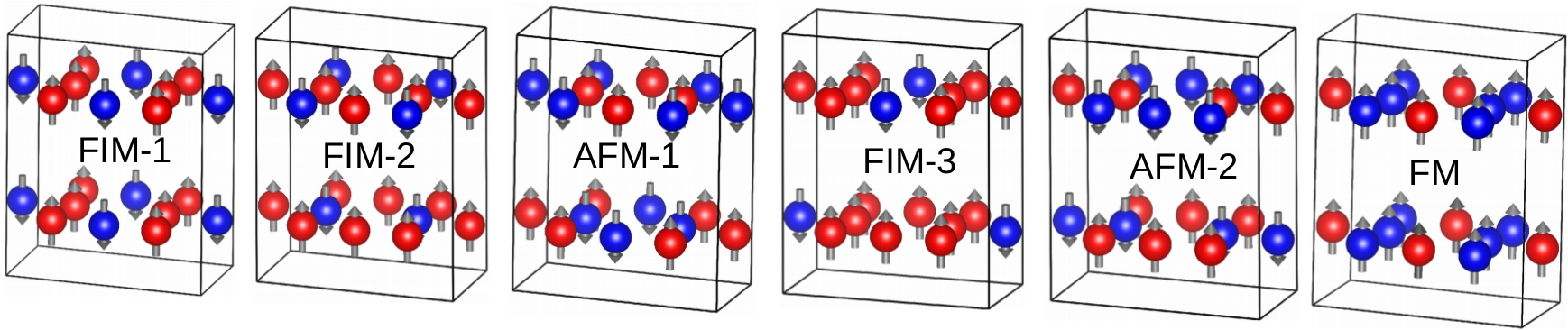,scale=0.24} 
  \caption{Considered spin configurations.
        From left to right, different ferrimagnetic (FIM),
        antiferromagnetic (AFM), and ferromagnetic (FM) arrangements
        are depicted in ascending order of their energy according to
        our calculations using experimental room-temperature lattice parameters. 
        Only Co atoms are shown. Different colors denote
        different spin orientations.  }
  \label{fig:mag}
\end{figure*}
Cs$_{2}$Co$_{3}$S$_4$ crystallizes in a body centered orthorhombic Bravais lattice with the centrosymmetric space group \textit{Ibam} (SG 72) and two formula units per primitive unit cell. Experimental lattice parameters and atomic positions were determined  both by XRD and by ND. While the ND data were obtained at a temperature of $4.2$ K, XRD was presumably performed at room temperature~\cite{bronger1988structure}. 

The experimental lattice parameters obtained by room-temperature XRD are $a$ = 5.712 \AA, $b$ = 11.231 \AA, and $c$ = 13.879 \AA. 
According to the same experiment, Cs and S atoms occupy the 8j and 16k sites at (0.2336, 0.1205, 0) and at (0.2233, 0.3669, 0.1589), respectively, while the Co atoms occupy the 8g sites, denoted as Co(I), at (0, 0.2319, 0.25) and the 4b sites, denoted as Co(II), at (0.5, 0, 0.25), see BHM.

The structure is shown in Fig. \ref{fig:crystal}. 
The Cs atoms are centered within distorted cubes of S atoms, and the Co atoms are surrounded by S tetrahedra. The related shortest bond lengths amount to 3.54 \AA{} for Cs - S, 2.31 \AA{} for Co(I) - S, and 2.34 \AA{} for Co(II) - S.

The experimental lattice parameters obtained by low-temperature ND
amount to $a$ = 5.663 \AA, $b$ = 11.160 \AA{} and $c$ = 13.708 \AA.
They are somewhat smaller than the related room-temperature parameters, in accordance with a positive thermal expansion coefficient. Atomic site positions obtained by ND are Cs-8j (0.234, 0.115, 0), S-16k (0.814, 0.369, 0.14), Co(I)-8g (0, 0.235, 0.25), and Co(II)-4b (0.5, 0, 0.25). These positions agree well with those determined by XRD with the exception of the sulfur position. The latter results in a Cs-S interatomic distance of $1.98$ \AA,
which is only somewhat more than half of the expected value of about $3.6$ \AA. Thus, we consider the ND-value of the S-16k atomic position as incorrect and use the XRD atomic positions in all DFT calculations.

The DFT calculations were performed with the full-potential local orbital code (FPLO),~\cite{PhysRevB.59.1743} version 18.00-52.
The standard generalized gradient approximation (GGA) based on 
the parameterization by Perdew, Burke, and Ernzerhof (PBE)~\cite{PhysRevLett.77.3865} was used for the  exchange-correlation potential. 
Local or semi-local approximations to DFT like GGA suffer from an incomplete description of orbital magnetism that
stems from orbital-dependent self-interaction.~\cite{Zhou09}
For this reason, so-called orbital polarization corrections (OPC) in the flavor suggested by Nordstr\"om {\em et al.}~\cite{Nordstrom92a} were applied in a part of the calculations.

With an aim to check the order of magnetic states, an additional set of calculations was performed with the WIEN2k code which is based on the full-potential linearized augmented plane wave method~\cite{blaha2001wien2k}.
For this, we use the Yukawa-screened (YS) hybrid functional YS-PBE0~\cite{tran2011implementation} with $\alpha = 0.25$ (Hartree-Fock exchange parameter), and $\lambda = 0.165$ (screening parameter) that give rise to results similar to the HSE06 functional.
Other parameters were chosen similar to FPLO calculations. 

A linear tetrahedron method with Bl\"ochl corrections was applied for the integrations in reciprocal space.
The related  $k$-mesh contained 12 $\times$ 12 $\times$ 12 points
in the Brillouin zone for the self-consistent calculations
and was refined to 20 $\times$ 20 $\times$ 20 points for the
evaluation of band structures and densities of states (DOS) shown 
in Fig.~\ref{fig:dosband}. 

Self-consistent calculations were conducted in both scalar relativistic 
and four component full relativistic modes of FPLO.
In the full relativistic mode that includes spin-orbit coupling in all orders, 
the direction of magnetization was fixed by a global setting of the spin quantization axis. 
Those calculations were Co atoms at the same Wyckoff position were
considered to carry magnetic moments with different orientation,
see Fig. \ref{fig:crystal}, were performed in space group {\em I222}
(SG 23). 
The fixed spin moment method (FSM)~\cite{Schwarz84} was applied to 
force a zero total spin moment for antiferromagnetic (AFM) arrangements. 

Maximally projected Wannier functions~\cite{Koe23} were produced from self-consistent full relativistic DFT band structures using the PYFPLO module of the FPLO code, version 18.00-52.
The localized Wannier basis consists of Cs [5p, 6s], Co [3d, 4s], and S [3p, 3s] orbitals, i.e., a total of 168 orbitals. To map the Wannier model,  the same $k$-mesh as in the self-consistent calculations was chosen. The generated Wannier Hamiltonians were used to interpolate the band  structure at a $k$-mesh of 40 $\times$ 40 $\times$ 40 points and to  search Weyl points with the help of the abovementioned PYFPLO module. The search was limited to the Wannier bands 129-131,  counted in ascending order, as lower bands at the Weyl node, see  Fig.~\ref{fig:dosband}(b). 

In the presence of spin-orbit interaction, nodal lines (NL) can only
exist on simple mirror planes~\cite{PhysRevB.92.085138}.
We confirmed the presence of NLs by calculating the Berry phase on a closed loop containing the NL.
See also the discussion of the gauge choice for link variables in the Appendix. 

The $k$-space integrations to evaluate the anomalous Hall conductivity were carried out with a mesh of 200 $\times$ 200 $\times$ 200 points.

\begin{table*}[tbh]
\caption{Scalar relativistic energies and spin moments.
Calculated relative total energies $E$, given per primitive unit cell with 18 atoms and
local and total spin magnetic moments
$M_s$ for the six considered configurations depicted in Fig.~\ref{fig:mag}. 
ND and XRD indicate the results obtained by using
low-temperature ND and room-temperature XRD lattice parameters, respectively.} 
\begin{tabular}{llllllllll}
\hline
              & \multicolumn{3}{c}{$E$ [meV/unit cell]~~} & \multicolumn{2}{c}{$M_s$(Co(I)) [$\mu_{\rm B}$]~~} & \multicolumn{2}{c}{$M_s$(Co(II)) [$\mu_{\rm B}$]~~} & \multicolumn{2}{c}{$M_s$(unit cell) [$\mu_{\rm B}$]} \\
\hline
Configuration~~~ & ND & XRD   & XRD    &  ND   & XRD   & ND   & XRD   & ND   & XRD          \\
XC-Functional & GGA~ & GGA~    & YS-PBE0~~~    &  GGA~   & GGA~   & GGA~   & GGA~   & GGA~   & GGA          \\
\hline
FIM-1 &         0    & 0     & 0     &  2.01 & 2.11 & -1.77 & -1.91 & 5.94 & 5.99 \\
FIM-2 &         19.6 & 20.4   & 37.5    &  1.98, -1.67~~  &2.09, -1.82~~~~ & 2.03, 2.05 &2.13, 2.16   & 5.96 & 6.00 \\
AFM-1 &         36.5 & 23.1    & 52.2   &  1.87, -1.87 &2.00, -2.00 & 2.00, -2.00~~ &2.14, -2.14~~~~ & 0    & 0 \\
FIM-3 & 	171  & 161    & 150    &  2.01, 2.04 & 2.17, -2.20& -1.71, 2.13&-1.80, 2.28& 10.87 & 11.58 \\
AFM-2 &         157  & 165  & 174      &  1.81, -1.81 &1.97, -1.97& 1.95, -1.95 &2.09, -2.09 & 0    & 0 \\
FM    &         367  & 388   & 370     &  1.92 &2.04& 1.92&2.04& 14.26 & 14.79 \\
\hline
\end{tabular}
\label{tab:mag}
\end{table*}

\begin{table*}[tbh]
\caption{Full relativistic energies, spin and orbital moments. Calculated relative total energies $E$, given per primitive unit cell with 18 atoms, and  local spin, orbital, and total magnetic moments ($M_s$, $M_l$, and $M_{\rm tot}$, resp.)
of the Co atoms. The two numbers in each column refer to Co(I) and Co(II). The upper part shows results of full relativistic calculations with different orientations of the magnetization for the configurations FIM-1 and AFM-1, the lower part shows results of full relativistic calculations with orbital polarization correction. Low-temperature ND lattice parameters were used for all calculations.}
\begin{tabular}{llllll}
\hline
With s-o        & & & & & \\
\hline
Configuration~~ & Direction of magnetization~~ & $E$ [meV/unit cell]~~ & $M_s$(Co) [$\mu_{\rm B}$]~~ & $M_l$(Co) [$\mu_{\rm B}$]~~ & $M_{\rm tot}$(Co) [$\mu_{\rm B}$] \\
\hline
FIM-1           & [1~0~0]                      & 0                     & 2.00, -1.76 & 0.17, -0.14 & 2.17, -1.90 \\
                & [0~1~0]                      & 2.7                   & 2.00, -1.77 & 0.15, -0.15 & 2.15, -1.92 \\
                & [0~0~1]                      & 2.5                   & 2.00, -1.76 & 0.16, -0.15 & 2.16, -1.92 \\
\hline
AFM-1           & [1~0~0]                      & 37.4                  & $\pm$ 1.86, $\pm$ 1.99~~~ & $\pm$ 0.15, $\pm$ 0.16~~~ & $\pm$ 2.01, $\pm$ 2.15 \\
                & [0~1~0]                      & 37.7                  & $\pm$ 1.86, $\pm$ 1.99 & $\pm$ 0.14, $\pm$ 0.16 & $\pm$ 2.00, $\pm$ 2.15 \\
                & [0~0~1]                      & 35.1                  & $\pm$ 1.86, $\pm$ 1.99 & $\pm$ 0.18, $\pm$ 0.18 & $\pm$ 2.04, $\pm$ 2.17 \\
\hline
\hline
With s-o and OPC   & & & & & \\
\hline
FIM-1           & [1~0~0]                      & 0                     & 1.99, -1.76 & 0.70, -0.47 & 2.69, -2.23 \\
                & [0~1~0]                      & 17.9                  & 2.01, -1.77 & 0.40, -0.56 & 2.41, -2.33 \\
                & [0~0~1]                      & 11.7                  & 2.01, -1.77 & 0.46, -0.58 & 2.47, -2.35 \\
\hline
AFM-1           & [1~0~0]                      & 46.3                  & $\pm$ 1.88, $\pm$ 2.01 & $\pm$ 0.50, $\pm$ 0.57 & $\pm$ 2.38, $\pm$ 2.58 \\
                & [0~1~0]                      & 51.7                  & $\pm$ 1.86, $\pm$ 1.99 & $\pm$ 0.44, $\pm$ 0.48 & $\pm$ 2.30, $\pm$ 2.47 \\
                & [0~0~1]                      & 12.6                  & $\pm$ 1.85, $\pm$ 1.96 & $\pm$ 0.71, $\pm$ 0.80 & $\pm$ 2.56, $\pm$ 2.76 \\
\hline
\end{tabular}
\label{tab:orbmag}
\end{table*}

\section{{RESULTS AND DISCUSSION}}

\subsection*{A. MAGNETIC GROUND STATE}
We started our calculations with a search for the stable magnetic
ground state of Cs$_2$Co$_3$S$_4$.

Based on ND data, a collinear antiferromagnetic (AFM) order was suggested 
in BHM
to be the stable magnetic state at 4.2 Kelvin. In this state, four of the Co(I) sites show a spin orientation opposite to the spin orientation of the four other Co(I) sites and two of the Co(II) sites show a spin orientation opposite to the spin orientation of the two other Co(II) sites (state AFM-1 in Fig.~\ref{fig:mag}). Permutation of the spin orientations of these four groups of sites yields eight different combinations. Two of the combinations are  equivalent by symmetry to other combinations. We investigated the six remaining combinations, shown in Fig.~\ref{fig:mag} which includes ferromagnetic (FM), two different AFM, and three ferrimagnetic (FIM)
arrangements. 

Table~\ref{tab:mag} compiles the relative total energies and spin magnetic moments, obtained by means of calculations in a scalar relativistic mode and using
both low-temperature ND and room-temperature XRD lattice parameters.
Different from the experimental result, the arrangement FIM-1 is found to be the lowest-energy state both within GGA and hybrid functional YS-PBE0~\cite{tran2011implementation}. However, both FIM-2 and the experimental ground state AFM-1 are further low-energy solutions with energies of  3 meV per atom or less above the computed ground state. All other considered arrangements have much higher total energies. 

In order to make sure that the obtained solutions with atomic spin moments of about 2 $\mu_{\rm B}$ are the only stable solutions of the given arrangements, we performed two tests. First, FSM calculations were done for the case of FM order. They showed that there is no other (meta)stable FM state in the possible range of local moments. Second, calculations for the AFM-1 order were performed with larger than usual initial spin splittings
of 4 and 5 $\mu_{\rm B}$. Both calculations converged to the same state as obtained with default initial splitting of 2 $\mu_{\rm B}$ and shown in Table~\ref{tab:mag}. 

The question arises why the experimental ground state AFM-1 is found higher in energy than the state FIM-1 in the calculations. To get an idea about a possible reason, we performed additional calculations for the states FIM-1 and AFM-1 with slight doping of 0.1 electrons (holes) per unit cell of 18 atoms. As a result, the original energy difference between FIM-1 and AFM-1 of 36.5 meV (ND lattice parameters) was reduced by 1.3 meV (enhanced by 1.4 meV). This means, a strong electron doping would be needed to change the order of the two states, which would only be possible by significant modification of the stoichiometry in the order of a few percent. 

We cannot rule out this possibility, nor can we rule out the applied approximation to the exchange-correlation functional (GGA) to be responsible for the disagreement between calculated and experimental ground states. In any case, the two discussed states are relatively close in energy. Thus, it should be possible to prepare FIM-1 either by appropriate modification of the stoichiometry (improved preparation or intentional electron doping) or by application of a moderate external magnetic field. The latter option would be
very interesting, since also the state FIM-2 is close to AFM-1.
Thus, a competition between three different phases could occur.

Next, we carried out full relativistic calculations 
including spin-orbit coupling. Here, we focus on the calculated ground state FIM-1 and on the experimental ground state AFM-1.
The results are presented in the upper part of Table~\ref{tab:orbmag}. 

By spin-orbit coupling, the total energy depends on the direction of magnetization. Here, we consider the three inequivalent orthorhombic directions [1 0 0], [0 1 0], and [0 0 1]. Among these, the lowest energy is found for the configuration FIM-1 with [1 0 0] orientation of the magnetization.
In the configuration AFM-1, the lowest energy is achieved if the
magnetization is oriented along [0 0 1]. This is in agreement
with the easy axis measured by ND in I. The maximum value of the magnetic anisotropy energy (MAE), which is the energy difference between unlike orientations of the magnetization,
amounts to about 2.5 meV per unit cell in both configurations.
The total magnetic moments per Co atom are not much enhanced in
comparison with the scalar relativistic results,
due to the relatively small orbital moments that do not exceed 10\%
of the spin moments. This finding is at variance with the experimental magnetic moments of about 3.7 $\mu_{\rm B}$, obtained by ND in I. We note, however, two important caveats.
First, the title system is a (half)metal with a relatively
broad Co-d band of about 3 eV, see next section.
As such, it is expected that the Co orbital moment is ruled
by itinerant behavior and, thus, much smaller than the large-moment
limit of the d$^7$ atomic configuration, 3 $\mu_{\rm B}$.
Second, there exists a generic problem to define the local
moment of atoms in an AFM configuration, since the total moment
of the unit cell has to be zero.  In the present AFM-1 configuration, Co atoms with antiparallel magnetic moments are less than 2.9 \AA{} away from each other, which is only 15\% more than the distance in densely packed metallic Co. Thus, it is possible to define very different local projections of the magnetization density which add up to the same total magnetization density.

Further, we applied so-called orbital polarization corrections (OPC)
to GGA. These corrections improve the compliance of GGA
and similar approximations to the exchange-correlation
functional of DFT with Hund's second rule.~\cite{Eschrig05}
Related results are compiled in the lower part of Table~\ref{tab:orbmag}. We note that the spin magnetic moments are merely unchanged in comparison with those obtained by scalar relativistic or full relativistic calculations without OPC. The orbital magnetic moments, however, are enhanced with respect to the data of the latter calculations by factors between
two and four. As a result, the total moments amount up to about 2.7 $\mu_{\rm B}$. This value is still significantly below the numbers reported, but much larger than cobalt moments commonly known from other metallic systems. Further, the lowest-energy directions of magnetization remain unchanged by OPC for the two considered configurations: [1 0 0] for the configuration FIM-1 and [0 0 1] for AFM-1. This fact is reassuring since AFM-1 [0 0 1] is the experimental ground state. Here, by applying GGA+OPC,
this state is found to be only less than
1 meV per atom higher in energy than the state FIM-1 [1 0 0].

We conclude this section by stating that we identify the state FIM-1 [1 0 0] as the ground state of the title compound, treated in relativistic GGA without or with OPC,
with exact stoichiometry and free of defects.
Since this state is very close in energy to the experimental ground state AFM-1 [0 0 1], we think that it can be obtained either directly by refined sample preparation or by application of a moderate external magnetic field. Thus, our further investigations will be focused on FIM-1.

\subsection*{B. ELECTRONIC STRUCTURE OF THE GROUND STATE}

\begin{figure*}[!htb]
\centering
\includegraphics[scale=0.335]{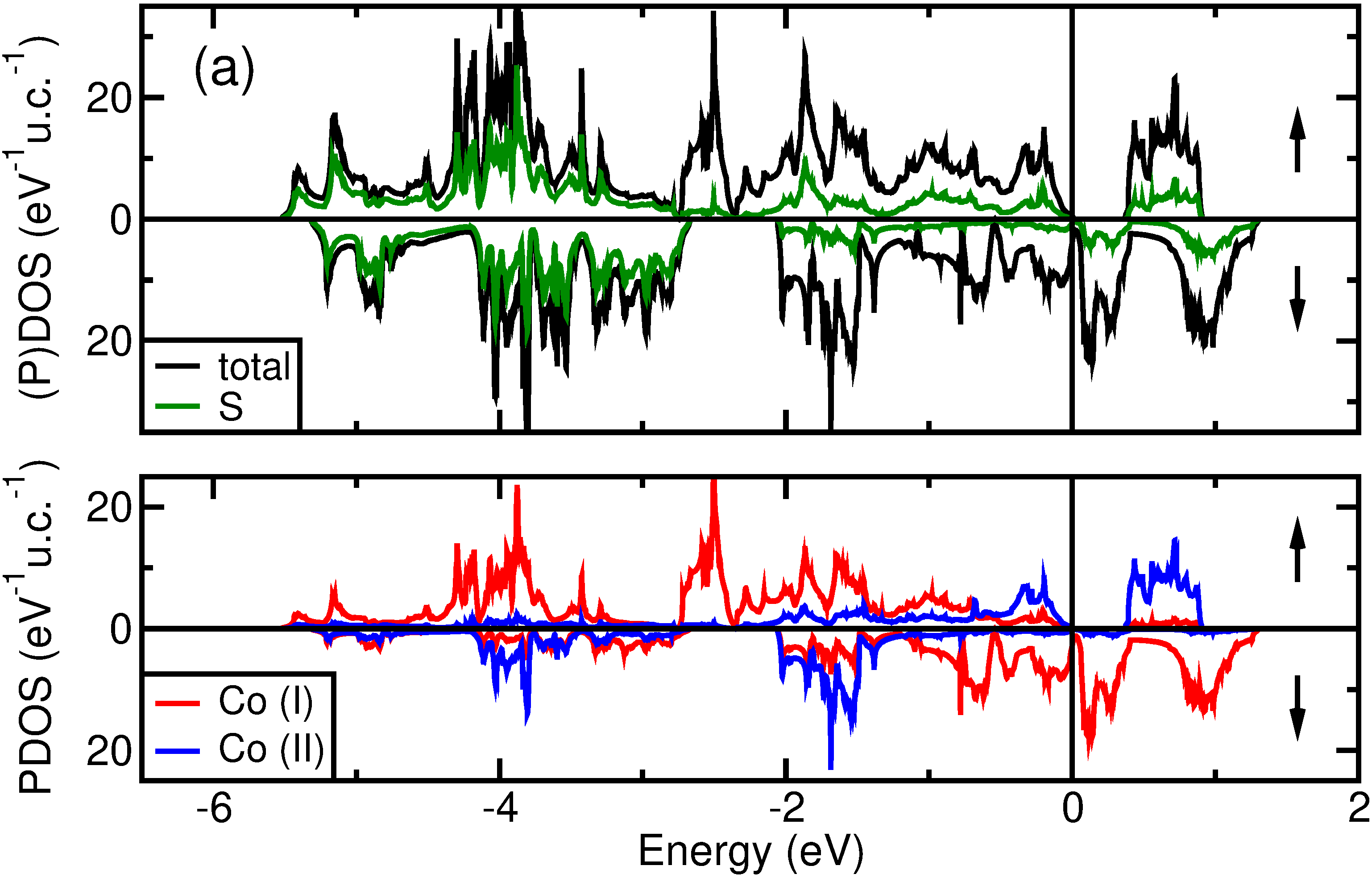}
\hspace{0.1cm}
\includegraphics[scale=0.355]{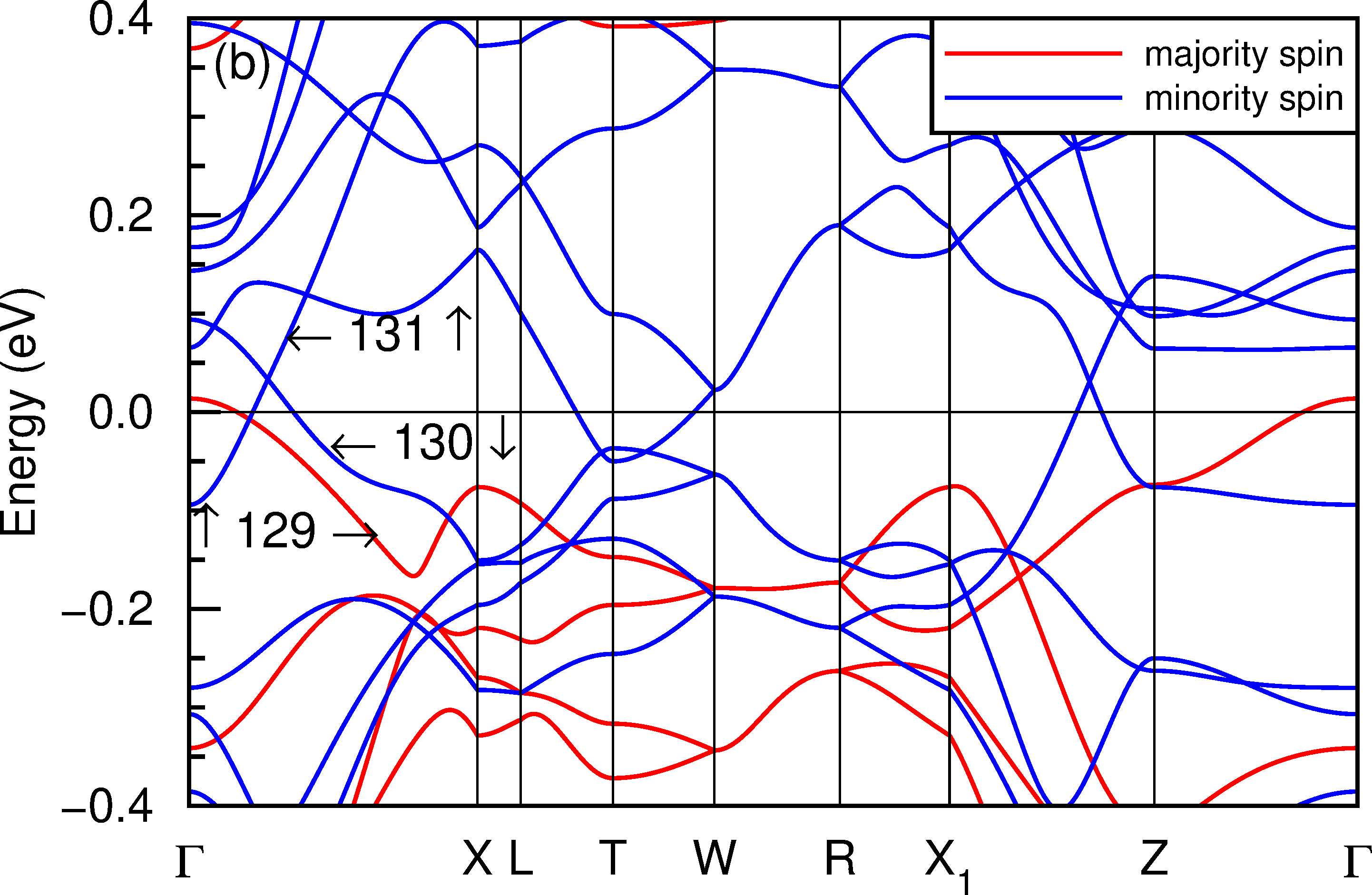}
\vspace{-0.39cm}
\includegraphics[scale=0.357]{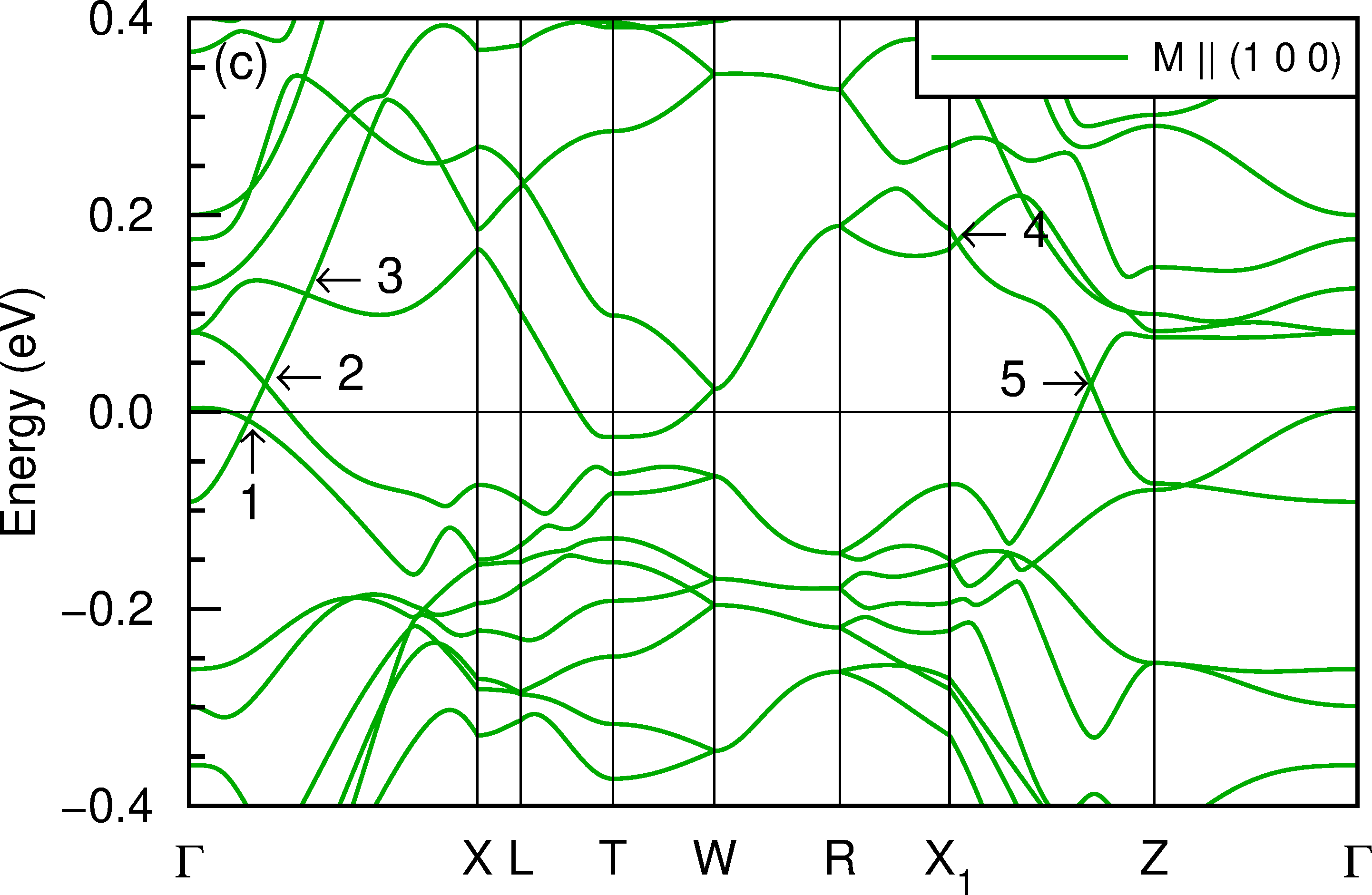}
\hspace{0.1cm}
\includegraphics[scale=0.357]{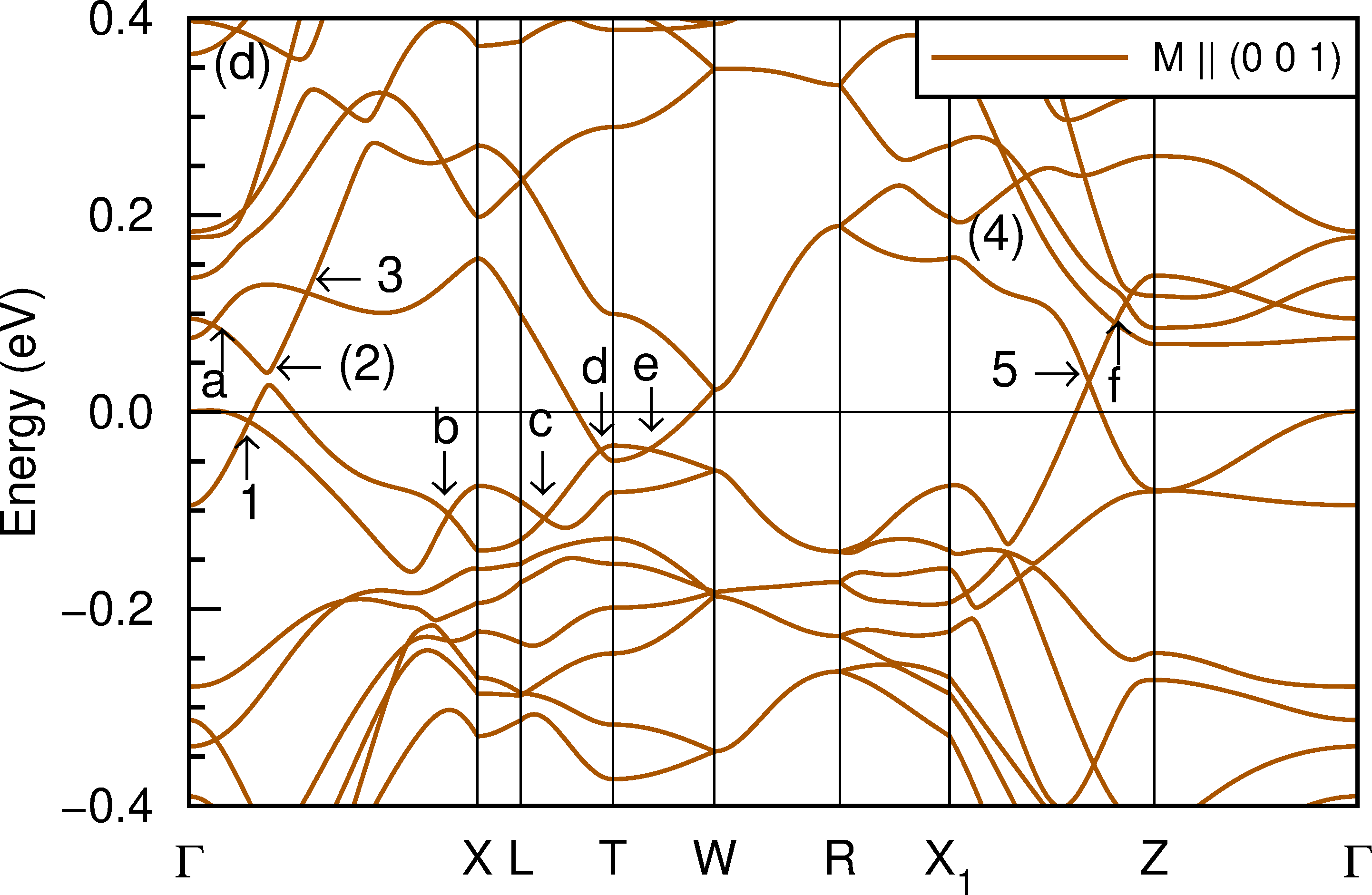}
\includegraphics[scale=0.355]{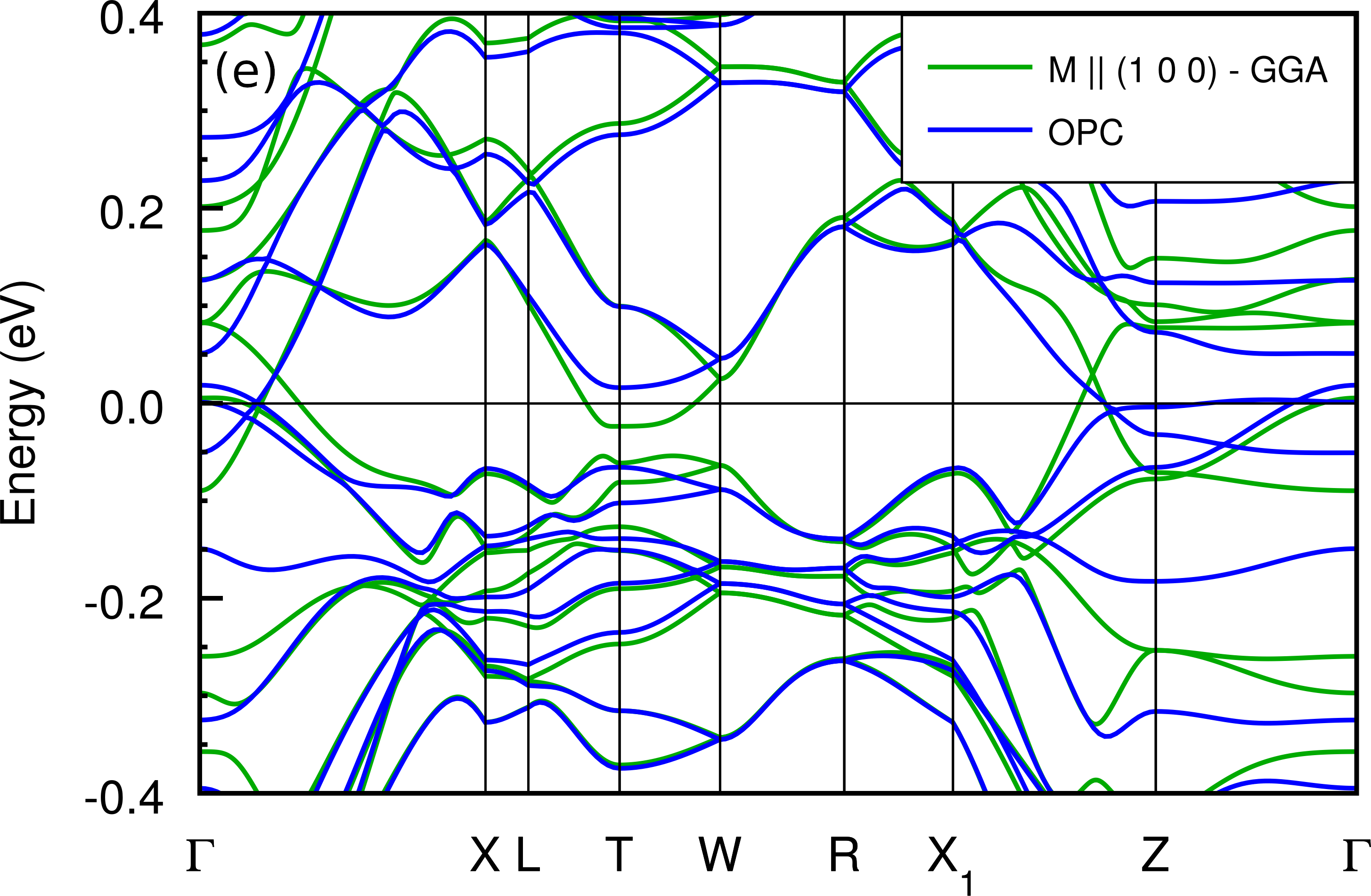}
\hspace{0.115cm}
\includegraphics[scale=0.355]{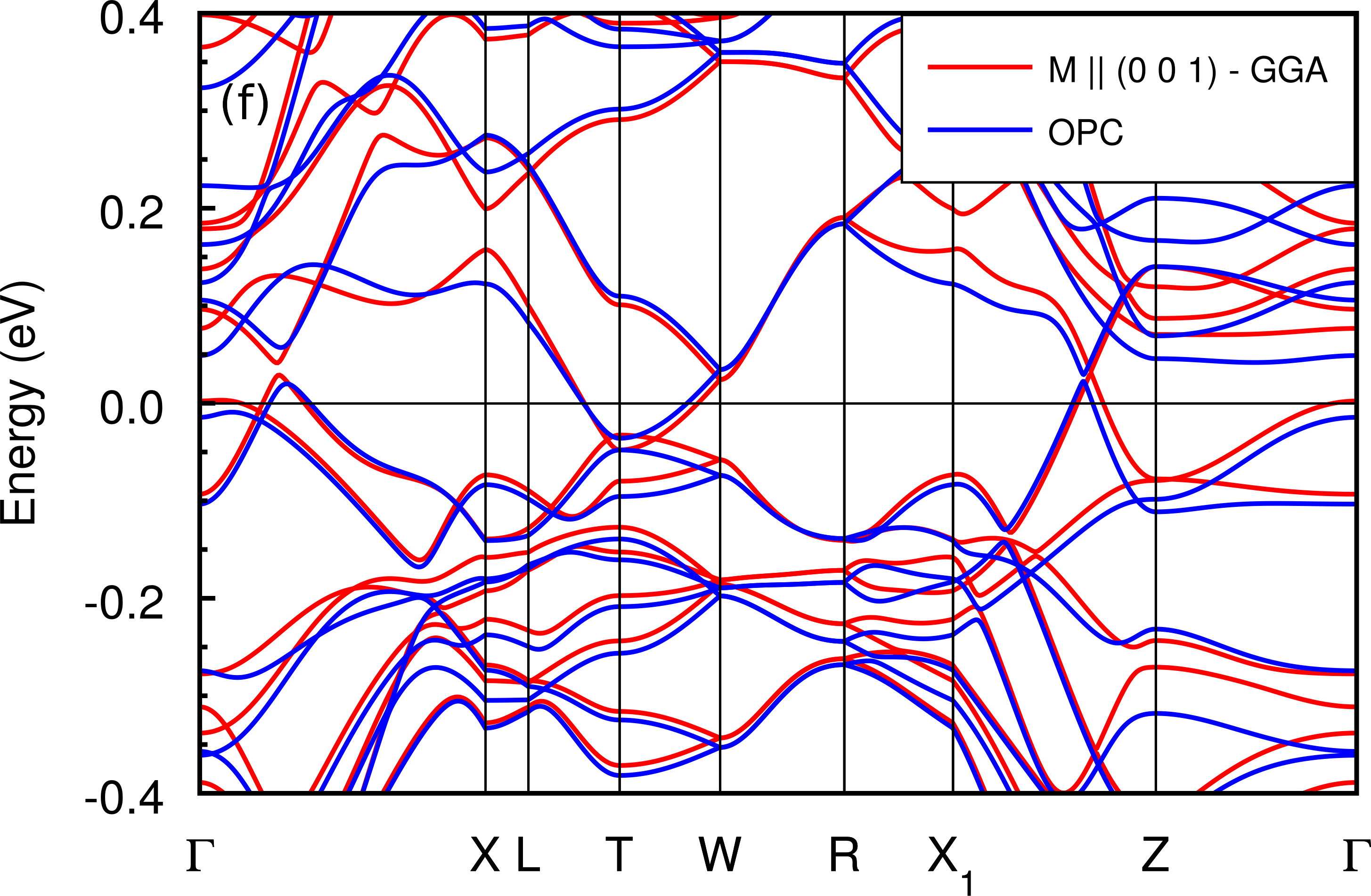}
\caption{
 Electronic structure of the ground state magnetic configuration.
The Fermi level is set to zero energy in all sub-figures and indicated
with solid lines. 
 (a) Total and partial DOS obtained in scalar relativistic mode. 
The majority and minority spin channels are indicated by up- and down-arrows,
respectively.
The total DOS is shown with black lines, the partial sulphur, Cobalt(I), and 
Cobalt(II) contributions with green, red, and blue lines, respectively.
(b) Electronic band structure obtained in scalar relativistic mode.
The majority and minority spin contributions are shown with red and blue
lines, respectively.
Several numbers of bands close to the Fermi level are given together with
arrows pointing to the related bands. Note, the numbering is in the order
of the band energy at each $k$-point.
(c) Electronic band structure obtained in full relativistic mode
with magnetization $M$ parallel to [1 0 0] (ground state).
Apparent band crossings involving bands $n$ and $n+1$, $n = \{129, 130, 131\}$,
are indicated with arrows and numbered.
(d) Electronic band structure obtained in full relativistic mode
with magnetization $M$ parallel to [0 0 1].
Apparent band crossings involving bands $n$ and $n+1$, $n = \{129, 130, 131\}$,
are indicated with arrows. Crossings already present for $M$ parallel to [1 0 0] 
 are numbered as before; former crossings that are obviously gapped are numbered
with parentheses; new crossings are distinguished by letters.
(e, f) Comparison between electronic band structures in
 full relativistic mode with magnetization $M$ parallel to [1 0 0] and [0 0 1], respectively,
 obtained with GGA (same data as in (c, d)) and with GGA+OPC. }

\label{fig:dosband}
\end{figure*}

All results of this and the following sections
were obtained by using room-temperature XRD structure data from I.

Fig.~\ref{fig:dosband} (a) shows the total and partial densities 
of states (PDOS) for the ground state FIM-1 configuration in 
scalar relativistic mode. 
Majority and minority spin channels are indicated with up- and
down-arrows, respectively.
The related scalar relativistic
band structure on high-symmetry lines is depicted 
in Fig.~\ref{fig:dosband} (b) and
the positions of the symmetry points are compiled in Table~\ref{tab:sympoints}.

\begin{table}
\label{tab:sympoints}
\caption{Location of the symmetry points used in Fig.~\ref{fig:dosband}.}
\begin{tabular}{cccc}
\hline
point   & $k_x [2 \pi / a]$ &   $k_y [2 \pi / b]$   & $k_z [2 \pi / c]$ \\
\hline
$\Gamma$ & 0                 &   0                   & 0               \\
X       & 0.58469           &   0                   & 0               \\
L       & 0.58469           &   0.17259             & 0               \\
T       & 1/2               &   1/2                 & 0               \\
W       & 1/2               &   1/2                 & 1/2             \\
R       & 1/2               &   0                   & 1/2             \\
X$_1$   & 0.41531           &   0                   & 1               \\
Z       & 0                 &   0                   & 1               \\
\hline
\end{tabular}
\end{table}

The DOS shows an interesting combination of semi-metallic
and covalent features. Sulphur states (green color in the upper
panel of Fig.~\ref{fig:dosband}(a)) dominate the
DOS between about -5 eV and about -3 eV (bonding region), while Co states
(lower panel of Fig.~\ref{fig:dosband}(a)) dominate between -2 eV 
and +1 eV (anti-bonding region).
Both regions are separated by a covalency gap in the minority
spin channel. In the majority spin channel, however, 
this energy range is occupied by states localized almost exclusively
at Co(I) sites. 

There is a narrow, separated band with a width of only 0.5 eV in the majority spin channel above the Fermi level.
However, the widths of the occupied majority spin band and 
of the minority spin band around the Fermi level in the Co-dominated anti-bonding region amount to about 3 eV. Thus, the title compound is expected to show itinerant magnetism. 

Within the majority spin band, there is a 0.36 eV gap only 14 meV above
the Fermi level, see Fig.~\ref{fig:dosband} (a, b). Thus, the system is very close to a half metallic state.
This finding is in accordance with the calculated spin moment
of 5.99 $\mu_{\rm B}$, which is almost an integer value that
would indicate half metallicity in a scalar relativistic approximation. Moreover, a narrow pseudogap of about 80 meV width is present in the minority spin band at $E_{\rm F}$, qualifying Cs$_2$Co$_3$S$_4$ as a semi-metal. The majority spin band has a small hole pocket at point $\Gamma$, see Fig.~\ref{fig:dosband} (b), and only two minority spin bands cross the Fermi level. Those bands which feature as highest occupied band in certain regions of the $k$-space are indicated in Fig.~\ref{fig:dosband} (b) with their band number (129, 130, or 131). Note that these numbers do not include the 120 semi-core states. 

The electronic band structure with spin-orbit coupling is shown in Fig.~\ref{fig:dosband} (c) for the case of magnetization along [1 0 0] (ground state).
Due to broken TRS, all bands are non-degenerate almost everywhere in the $k$-space. Exceptions are bands on the line W-R, which are at least two-fold degenerate by symmetry, and accidental band degeneracies. The latter may form so-called nodal lines or are isolated Weyl points, which are of special interest for their impact on transport behavior, see below. 

Inspection of Fig.~\ref{fig:dosband} (c) shows a number of 
apparent band crossings. We call these degeneracies ``apparent'', as the information provided by Fig.~\ref{fig:dosband} (c) does not allow a final categorization into Weyl points, nodal lines, or gapped states. Those apparent crossings that take place between bands $n$ and $n+1$ close to the Fermi level, $n = \{129, 130, 131\}$, are marked in Fig.~\ref{fig:dosband} (c) by numbers
1 to 5 and will be investigated in the following section.

If the magnetization is rotated by, e.g., application of an external
magnetic field, the band structure changes due to spin-orbit coupling. This means that also the positions and energies of Weyl points depend on the direction of magnetization, and pairs of Weyl points may even annihilate or emerge.~\cite{PhysRevResearch.1.032044}
This effect becomes obvious by comparing  Fig.~\ref{fig:dosband} (c)
with Fig.~\ref{fig:dosband} (d). The latter figure shows the band structure for magnetization along [0 0 1]. Out of the five apparent band crossings marked in Fig.~\ref{fig:dosband} (c), two are now visibly split. Those, (2) and (4), are marked in parentheses in Fig.~\ref{fig:dosband} (d). Moreover, six new apparent crossings manifest themselves, marked with letters a to f.

Fig.~\ref{fig:dosband} (e, f) show comparisons between GGA and GGA+OPC band structures.
OPC does in general not change the symmetry of bands that are already 
subject to spin-orbit and exchange splittings.
However, it can considerably enhance the existing splittings in proportion to the local 
orbital moments. Here, related band shifts up to several 10 meV are observed in the vicinity
of the Fermi level. The effect of OPC on band crossings is of quantitative nature for 
$M ||$ [1 0 0], Fig.~\ref{fig:dosband} (e), but disappearance of the degeneracies d and e
by the action of OPC is observed for $M ||$ [0 0 1], Fig.~\ref{fig:dosband} (f).
Below, we will study the effect of OPC on the target quantity of this work, the AHC.

\subsection*{C. WEYL POINTS}
We are now going to clarify the character of the apparent band crossings found in the previous section and to investigate their dependence on the direction of magnetization. For this aim, sets of Weyl points were searched using the Wannier presentation of the DFT band structure among all bands $n$ and $n+1$, $n = \{129, 130, 131\}$. For comparison, or if the visible crossing was not detected as a Weyl point while using the search algorithm, additional checks using the DFT band structure were performed.

Tab.~\ref{tab:points0} shows the results for the ground state with $M ||$ [1 0 0]. For this orientation of $M$, the magnetic point group includes a simple  (i.e., without time reversal) $C_2(x)$ rotation axis. All five apparent crossings lie on this axis, including points 4 and 5 which can be shifted to a position with $k_y = k_z = 0$ by subtracting a reciprocal lattice vector.
Along any simple  $C_2$ axis, crossings between bands belonging to the two different irreps are always WPs.~\cite{PhysRevB.92.085138,PhysRevLett.108.266802}
All five points are indeed confirmed as WPs with chiralities $\chi$ of 1 or -1. Their energies and $k_x$-positions obtained from the Wannier Hamiltonian are close to the related DFT results.

\begin{table}[!h ]
  \centering
  \caption{Characteristics of points in the low-energy electronic structure of Cs$_2$Co$_3$S$_4$, if the magnetization points along the [1 0 0] direction. ``P'' denotes the specific points as defined in Fig.~\ref{fig:dosband} (c); $E$ is the energy relative to the Fermi level; $k_i$, $(i = x, y, z)$ are the coordinates of one representative position of the point set; ``M'' is the symmetry-determined multiplicity of the positions; ``T'' is the type of the node (WP: confirmed Weyl point); $\chi$ is the chirality. Energies and positions obtained from the DFT calculation are given in parentheses.}
  \begin{ruledtabular}
  \begin{tabular}{c c c c c c c c}
   P & E[meV] & $k_x [2 \pi / a]$ & $k_y [2 \pi / b]$ & $k_z [2 \pi / c]$ & M & T & $\chi$\\\cline{1-8}
   1 & -6     & 0.118             & 0                 & 0                 & 2 & WP& 1     \\
     & (-8)   & (0.119)           & (0)               & (0)               &   &   &       \\\cline{1-8}
     
   2 & 32     & 0.153             & 0                 & 0                 & 2 & WP& -1    \\
     & (28)   & (0.154)           & (0)               & (0)               &   &   &       \\\cline{1-8}
     
3 & 123     & 0.236             & 0                 & 0                 & 2 & WP& 1    \\   
    & (120)  & (0.238)           & (0)               & (0)               & 2 &WP&      \\\cline{1-8}
    
4 & 179     & 0.399             & 0                 & 1                 & 2 & WP& -1    \\       
    & (173)  & (0.400)           & (0)               & (1)               & 2 &WP&       \\\cline{1-8}  
   
   5 & 33     & 0.132             & 0                 & 1                 & 2 & WP& 1     \\
     & (28)   & (0.129)           & (0)               & (1)               &   &   &       \\
  \end{tabular}
  \end{ruledtabular}
  \label{tab:points0}
\end{table} 

If the magnetization is rotated away from the easy [1 0 0] axis,
the electronic structure changes.
As a consequence, the Weyl points move both in energy and 
position.~\cite{PhysRevB.92.085138,PhysRevResearch.1.032044}
Tab.~\ref{tab:point2} and Tab.~\ref{tab:point5} show, how the angle $\phi$ of magnetization, varied in steps of 15 degrees from [1 0 0] toward [0 0 1], affects point 2 and point 5, respectively.
We note, that the band energy $E$ and $k_x$ are only marginally shifted. However, the WPs move strongly in the direction of $k_z$ in both cases. For $M ||$ [$M_x$ 0 $M_z$], $M_z \neq 0$,
the $m'_z$ plane (mirror plane with time reversal)
is not contained in the magnetic point group, in distinction to $M ||$ [1 0 0]. Thus, for $\phi \neq 0$, the multiplicity of the WPs does not depend on $k_z$. Especially, $k_z = 0$ is not a plane with a symmetry different from $k_z \neq 0$.
Vice versa, a set of WPs with multiplicity 2 and coordinates ($k_x$, 0, $k_z \neq 0$)
for the case of $M ||$ [$M_x$ 0 $M_z$], $M_z \neq 0$, has to move toward the plane $k_z = 0$ if $M$ rotates toward [1 0 0] in order to keep its multiplicity. 

If $\phi$ exceeds 45 degrees, none of the WP pairs can be detected. Obviously, they are annihilated before $\phi$ reaches the value of 60 degrees.

\begin{table}[h!]
  \centering
  \caption{Impact of magnetization direction on the WP number 2.
  $\phi$ denotes the deflection of the magnetization direction from [1 0 0] toward [0 0 1], in degree; $E$ is the energy relative to the Fermi level; $k_i$, $(i = x, y, z)$ are the coordinates of one representative position of the WP set;
 ``M'' is the symmetry-determined multiplicity of the positions. 
 Note, the data for $\phi =0$ are the same as given in Tab.~\ref{tab:points0}}.
 \begin{ruledtabular}
  \begin{tabular}{c c c c c c}
   $\phi$ & E[meV] & $k_x [2 \pi / a]$ & $k_y [2 \pi / b]$ & $k_z [2 \pi / c]$ & M \\\cline{1-6}
   0      & 32     & 0.153             & 0                 & 0                 & 2 \\ 
   15     & 31     & 0.153             & 0                 & 0.076             & 2 \\ 
   30     & 31     & 0.154             & 0                 & 0.179             & 2 \\
   45     & 30     & 0.155             & 0                 & 0.339             & 2 \\
  \end{tabular}
  \end{ruledtabular}
  \label{tab:point2}
\end{table}

\begin{table}[h!]
  \centering
  \caption{Impact of magnetization direction on the WP number 5.
  $\phi$ denotes the deflection of the magnetization direction from [1 0 0] toward [0 0 1], in degree; $E$ is the energy relative to the Fermi level; $k_i$, $(i = x, y, z)$ are the coordinates of one representative position of the WP set;
 ``M'' is the symmetry-determined multiplicity of the positions.
 Note, the data for $\phi =0$ are the same as given in Tab.~\ref{tab:points0}}
 \begin{ruledtabular}
  \begin{tabular}{c c c c c c}
   $\phi$ & E[meV] & $k_x [2 \pi / a]$ & $k_y [2 \pi / b]$ & $k_z [2 \pi / c]$ & M \\\cline{1-6}
    0     & 33     & 0.132             & 0                 & 1                 & 2 \\ 
   15     & 31     & 0.134             & 0                 & 0.898             & 2 \\ 
   30     & 31     & 0.135             & 0                 & 0.851             & 2 \\
   45     & 30     & 0.142             & 0                 & 0.691             & 2 \\
  \end{tabular}
  \end{ruledtabular}
  \label{tab:point5}
\end{table}

The magnetic point group changes again, if the magnetization direction reaches [0 0 1].
In this case, $m(z)$ becomes a simple mirror plane, and the same 
holds for the equivalent plane $k_z = 2\pi /c$. 
Such a plane cannot host WPs, since the chiral charge is odd under reflection, 
and the generic degeneracies are nodal lines.~\cite{PhysRevB.92.085138} 
Indeed, we found only one WP among the apparent degeneracies (point e), 
see Tab.~\ref{tab:points90}. 
Another WP (point g, which is not on a symmetry line) 
was found during the general search. 
As required by symmetry, the two WPs e and g are not situated on a mirror plane.
Two of the other apparent degeneracies were found to be gapped (point 3, point 5).
The remaining six degeneracies, which all are located on a mirror plane, were identified
as points on a nodal line.
A comparison between the Wannier data and the DFT results for the 
degenerate points shows,
that the related band energies almost agree. 
This holds as well for the $k_x$ and $k_y$ coordinates, 
while the values for $k_z$ deviate somewhat stronger from each other.

\begin{table}[!h ]

  \centering 
    \caption{Characteristics of points in the low-energy electronic structure of Cs$_2$Co$_3$S$_4$, 
if the magnetization points along the [0 0 1] direction. 
 ``P'' denotes the specific points as defined in Fig.~\ref{fig:dosband} (d) and one more
 point, g, which is not situated on a symmetry line;
 $E$ is the energy relative to the Fermi level;
 $k_i$, $(i = x, y, z)$ are the coordinates of one representative position of the point set;
 ``M'' is the symmetry-determined multiplicity of the positions;
 ``T'' is the type of the point (WP: confirmed Weyl point;
 NL: a point on a nodal line; G: gapped);
 $\chi$ is the chirality in case of a confirmed Weyl point.
 Energies and positions obtained from the DFT calculation are given in parentheses.
 }

  \begin{ruledtabular}
  \begin{tabular}{c c c c c c c c}
   P & E[meV] & $k_x [2 \pi / a]$ & $k_y [2 \pi / b]$ & $k_z [2 \pi / c]$ & M & T & $\chi$\\\cline{1-8}
   a &  92    & 0.043             & 0                 & 0                 & 2 & NL &     \\
     & (89)   & (0.047)           & (0)               & (0)               &   &    &     \\\cline{1-8}
   1 &  -9    & 0.120             & 0                 & 0                 & 2 & NL &     \\
     & (-10)  & (0.121)           & (0)               & (0)               &   &    &     \\\cline{1-8}
   3 & 123    & 0.237             & 0                 & 0                 & 2 & G  &     \\
     & (121)  & (0.239)           & (0)               & (0)               &   &    &     \\\cline{1-8}
   b & -104   & 0.527             & 0                 & 0                 & 2 & NL &     \\
     & (-104) & (0.526)           & (0)               & (0)               &   &    &     \\\cline{1-8}
   c & -109   & 0.563             & 0.254             & 0                 & 4 & NL &     \\
     & (-108) & (0.563)           & (0.255)           & (0)               &   &    &     \\\cline{1-8}
   d & -38    & 0.510             & 0.460             & 0                 & 4 & NL &     \\
     & (-39)  & (0.511)           & (0.456)           & (0)               &   &    &     \\\cline{1-8}
   e & -36    & 0.5               & 0.5               & 0.140             & 4 & WP & -1  \\
     & (-38)  & (0.5)		  & (0.5)             & (0.176)           &   &    &     \\\cline{1-8}
   5 & 34     & 0.136             & 0                 & 1                 & 2 & G  &     \\
     & (31)   & (0.133)           & (0)               & (1)               &   &    &     \\\cline{1-8}
   f & 91     & 0.084             & 0                 & 1                 & 2 & NL &     \\ 
     & (90)   & (0.080)           & (0)               & (1)               &   &    &     \\\cline{1-8}
   g & 35     & 0.863             & 0                 & 0.112             & 4 & WP & -1   \\
     & (35)   & (0.865)           & (0)               & (0.141)           &   &    &     \\
  \end{tabular}
  \end{ruledtabular}
  \label{tab:points90}
 \end{table} 

\subsection*{D. ANOMALOUS HALL CONDUCTIVITY}

The anomalous Hall effect (AHE) is a notable characteristic of magnetic materials. Three particular contributions to the anomalous Hall conductivity (AHC), $\sigma^{\rm an}_{xy}$ are commonly distinguished: the intrinsic contribution, $\sigma^{\rm an, in}$, which arises solely from the material's band structure,
the skew scattering contribution $\sigma^{\rm an, sk}$,
and the side jump contribution $\sigma^{\rm an, sj}$.
The latter two arise from extrinsic mechanisms.~\cite{RevModPhys.82.1539} All three contributions are additive, $\sigma^{\rm an} =
\sigma^{\rm an, in} + \sigma^{\rm an, sk} + \sigma^{\rm an, sj}$.
Here, we consider only the intrinsic contribution to the AHC, which 
is known to reach large values in magnetic WSMs.~\cite{liu2018giant,PhysRevX.8.041045,PhysRevB.103.144410,PhysRevB.105.035107,PhysRevB.97.060406}
This behavior is considered to be due to the non-trivial band topology at  WPs with related large Berry curvature $\Omega(k)$.
As $k-$space analogy to magnetic dipoles, pairs of WPs act as 
sources and sinks of Berry curvature which, in turn, generates the AHC. 

The Kubo-formula linear-response approach was used to evaluate the AHC from the Berry curvature,~\cite{PhysRevB.103.205104}
based on the available tight-binding Hamiltonian:

\begin{equation}
 \sigma^{\rm an, in}_{ab}=-\frac{e^2}{\hbar}\Sigma_{n,c}
 \epsilon_{abc}\int_{BZ}\frac{d^3k}{(2\pi)^3}f_0(E_n(k))\Omega^c_{n}(k),
\end{equation}
where $\epsilon_{abc}$ is the Levi-Civita tensor,
$f_0(E_n(k))$ is the equilibrium Fermi-Dirac distribution function, 
n is the band index, $E_n(k)$ is the related eigenvalue at $k$,
and $\Omega^c_{n}$ is the $c$-component of the Berry curvature.

\begin{figure}[!htb]
\centering
\includegraphics[scale=0.34]{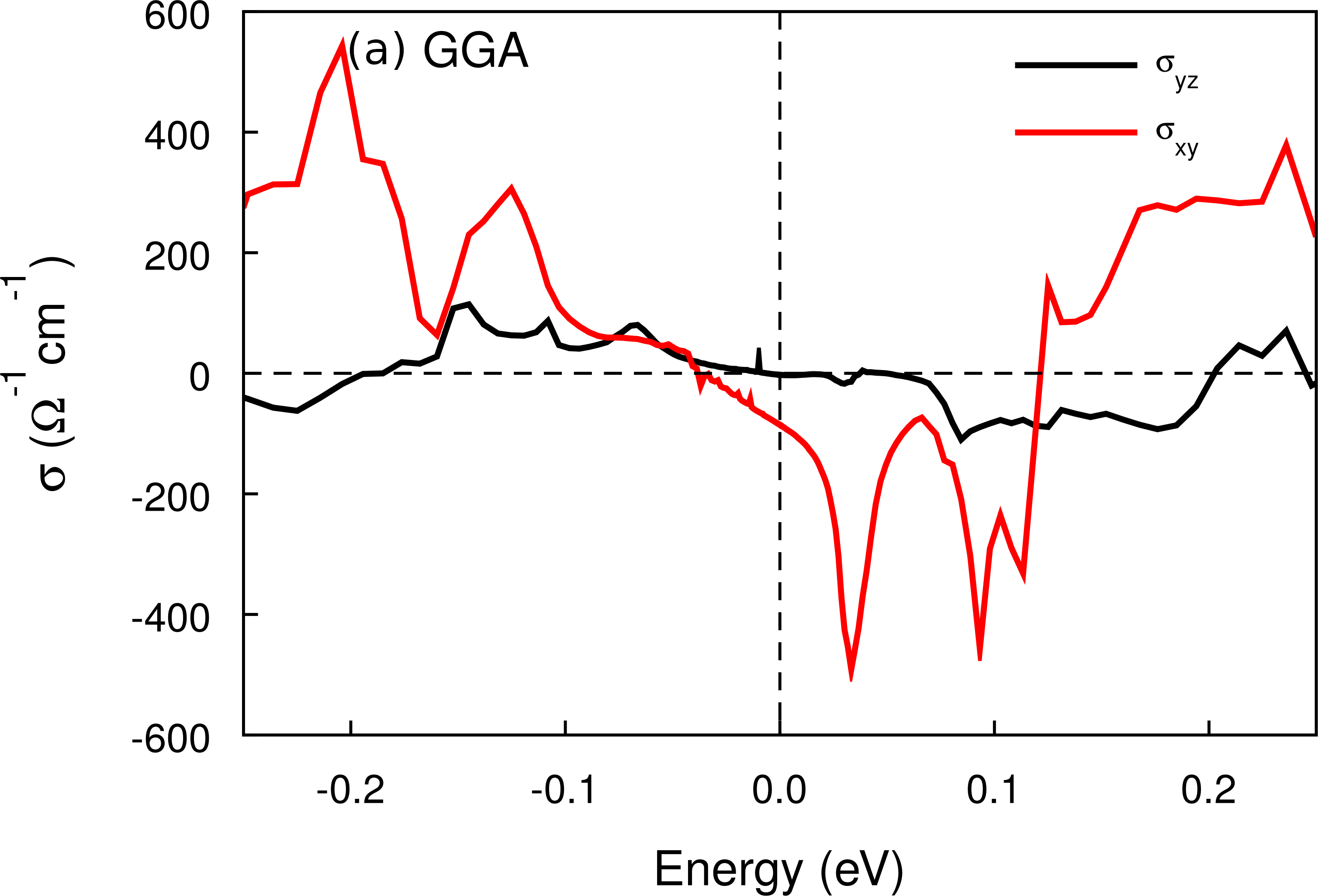}\\
\vspace{-0.85cm}
\includegraphics[scale=0.34]{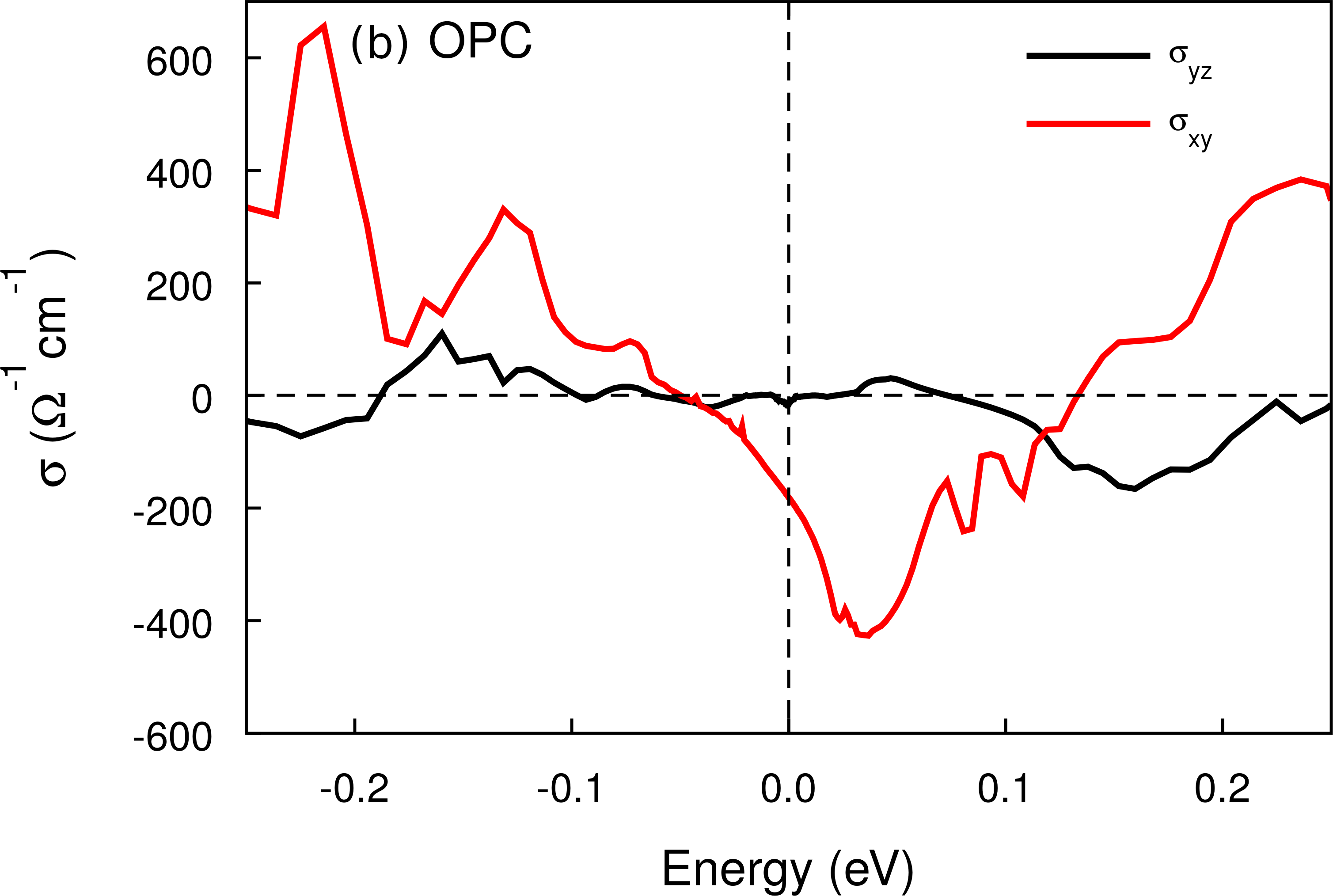}\\
\vspace{-0.2cm}
\caption{Intrinsic anomalous Hall conductivity of
Cs$_2$Co$_3$S$_4$ as a function of the chemical potential.
The AHC $\sigma_{yz}$ and $\sigma_{xy}$ were obtained for
magnetization directions along [100] and [001], within (a) GGA and (b) GGA+OPC, respectively.} 
\label{fig:ahc}
\end{figure}

Fig.~\ref{fig:ahc} (a, b) shows the band-intrinsic contribution to the
AHC as a function of the chemical potential and the
magnetization direction, obtained with a zero-temperature
Fermi-Dirac distribution function, computed within GGA and GGA+OPC functionals.
For the two considered magnetization directions [100] and [001],
$\sigma^{\rm an, in}_{yz}$ and $\sigma^{\rm an, in}_{xy}$, respectively, were evaluated. 
Roughly, the shape and magnitude of AHC is similar for the two 
approximations used. A closer inspection shows that the AHC obtained with GGA
is more structured than that one obtained with GGA+OPC. We attribute this difference to
the additional $k$-dependent band splittings produced by OPC.

In the ground state with easy magnetization axis [100], 
the calculated AHC is found to be nearly zero at $E_{\textrm F}$, 
but finite with $\sim$ -15 (27) $\Omega^{-1}$cm$^{-1}$ around 30 meV above
$E_{\textrm F}$ within GGA (GGA+OPC), where four WPs were found. 
However, this feature is much less prominent than the extended regions of large $|\sigma^{\rm an, in}_{yz}|$ farther away from the Fermi level. These regions may originate from WPs that were not found in our limited search described in the previous section. 
However, it is known that in general the 
combination of broken time reversal symmetry with spin-orbit
coupling generates contributions to the AHC also for bands with 
zero Chern number.~\cite{RevModPhys.82.1959}
Thus, the relation between AHC and Weyl points
can be relatively weak, as already observed, e.g., 
in Ref.~\onlinecite{PhysRevResearch.1.032044}.

For the [001] direction of magnetization, on the other hand, the
calculated AHC at $E_{\textrm F}$ amounts to 
-86 (-184) $\Omega^{-1}$cm$^{-1}$ within GGA (GGA+OPC). 
At around 30 meV above $E_{\textrm F}$, 
a prominent AHC peak is observed  with a magnitude of 
$\sim 500$ (430) $\Omega^{-1}$cm$^{-1}$, appearing at the 
position of the observed WP (point g).
It is worth noting that, with a small amount of electron doping, 
a large value of AHC can be attained at $E_{\textrm F}$.

\section{Summary}
To summarize, we investigated the electronic structure, the magnetic properties, and the intrinsic contributions to the anomalous Hall effect of Cs$_{2}$Co$_{3}$S$_4$ by means of density-functional theory calculations. We found a ferrimagnetic ground state of the title compound, at variance with the original experimental results that report an antiferromagnetic low-temperature state. The energy difference between the two states is in the order of meV/atom. Thus, a very small of-stoichiometry of the samples could have influenced the experimental result; likewise, one should keep in mind the limitations of the chosen approximations to the density functional. 

We characterized the theoretical ground state as a semi-metal that is very close to a half-metallic state. Several Weyl points are situated close to the Fermi level for the lowest-energy orientation of magnetization along [1 0 0]. They move both in $k$ and energy if the magnetization is rotated and are eventually annihilated. Related, we found the intrinsic anomalous Hall conductivity to be very sensitive to the orientation of the magnetization. It is relatively small for the easy-axis orientation but reaches magnitudes up to 500 $\Omega^{-1}$cm$^{-1}$ if the magnetization is oriented along [0 0 1]. 

Further experimental efforts on this interesting, though scarcely investigated
system are encouraged.

\section*{ACKNOWLEDGMENTS}
This work was supported by a grant from UNESCO-TWAS and the Swedish International Development Cooperation Agency (SIDA) (award no. 21-377 RG/PHYS/AS\textunderscore G). The views expressed herein do not necessarily represent those of UNESCO-TWAS, SIDA or its Board of Governors. M.P.G. acknowledges the Alexander von Humboldt
Foundation, Germany for the equipment grants and
IFW-Dresden for providing large-scale compute
nodes to Tribhuvan University for
scientific computations. 
M.P.G. thanks the University Grants Commission, Nepal for the Collaborative Research Grants (award no. CRG-78/79 S\&T-03). 
 G.B.A. thanks Nepal Academy of Science and Technology for the PhD fellowship.
All authors thank Ulrike Nitzsche for her skillful technical assistance.

\appendix
\section{Gauge choice for link variables}

The Berry phase $\gamma=\oint\boldsymbol{A}d\boldsymbol{k}$ collected
along a closed loop encircling a nodal line is an odd multiple of
$\pi$. The well known discretization\cite{King1993} $\gamma=i\sum_{i=0}^{N-1}\ln\left\langle u^{\boldsymbol{k}_{i}}\mid u^{\boldsymbol{k}_{i+1\mod N}}\right\rangle $
of this integral ($u^{\boldsymbol{k}}$ is the periodic part of the
wave function $\Psi^{\boldsymbol{k}}$) contains the link variable
$U^{\boldsymbol{k}_{i}}=\left\langle u^{\boldsymbol{k}_{i}}\mid u^{\boldsymbol{k}_{i+1\mod N}}\right\rangle $,
which can be generalized to the multi band case~\cite{Fuk2005} to
safely handle degeneracies. 

We used such integrals based on an FPLO Wannier model to prove the
existence of nodal lines in our compound.

To complement the discussion in Ref.~\onlinecite{Koe23} of the proper approximation
to a neglected position operator contribution to the Berry connection we will show the corresponding choice for link variables.

For a Wannier model one has wave function $\Psi_{n}^{\boldsymbol{k}}=\left(\Phi^{\boldsymbol{k}}C^{\boldsymbol{k}}\right)_{n}=\sum_{\boldsymbol{s}\mu}\Phi_{\boldsymbol{s}\mu}^{\boldsymbol{k}}C_{\boldsymbol{s}\mu,n}^{\boldsymbol{k}}$
of band $n$ with Bloch sums $\Phi_{\boldsymbol{s}\mu}^{\boldsymbol{k}}$
of Wannier functions at site $\boldsymbol{s}$ with quantum numbers $\mu$.
The Berry connection matrix then has two terms, one containing the
basis connection matrix $\boldsymbol{A}_{\Phi}^{\boldsymbol{k}}=\left\langle \Phi^{\boldsymbol{k}+}\mid\boldsymbol{\beta}_{\boldsymbol{k}}\Phi^{\boldsymbol{k}}\right\rangle $
(with Berry operator $\boldsymbol{\beta}_{\boldsymbol{k}}=\mathrm{e}^{i\boldsymbol{k}\boldsymbol{r}}i\boldsymbol{\nabla}_{\boldsymbol{k}}\mathrm{e}^{-i\boldsymbol{k}\boldsymbol{r}}$)
and a coefficient term
\[
\boldsymbol{A}_{\Psi}^{\boldsymbol{k}}=C^{\boldsymbol{k}+}\boldsymbol{A}_{\Phi}^{\boldsymbol{k}}C^{\boldsymbol{k}}+C^{\boldsymbol{k}+}S^{\boldsymbol{k}}i\boldsymbol{\nabla}_{\boldsymbol{k}}C^{\boldsymbol{k}}
\]
where $S^{\boldsymbol{k}}=\left\langle \Phi^{\boldsymbol{k}}\mid\Phi^{\boldsymbol{k}}\right\rangle $
is the overlap matrix (for generality). While the basis connection
$\boldsymbol{A}_{\Phi}^{\boldsymbol{k}}$ is related to the position
operator in Wannier basis, in models one does not know that term.
Hence, for simplification or for model situations an approximation
of $\boldsymbol{A}_{\Phi}^{\boldsymbol{k}}$ is needed. It was shown
in Ref.~\onlinecite{Koe23} that a simple approximation $\boldsymbol{A}_{\Phi,\boldsymbol{s}^{\prime}\boldsymbol{s}}^{\boldsymbol{k}}=\overline{\lambda}S_{\boldsymbol{s}^{\prime}\boldsymbol{s}}^{\boldsymbol{k}}\boldsymbol{s}$
(with $\boldsymbol{s}$ and $\boldsymbol{s}^{\prime}$ being site
vectors) yields proper crystal symmetry of all results. It depends
on the choice of the gauge of the Bloch sums $\Phi_{\boldsymbol{s}}^{\boldsymbol{k}\lambda}\left(\boldsymbol{r}\right)=\frac{1}{\sqrt{N}}\sum_{\boldsymbol{R}}\mathrm{e}^{i\boldsymbol{k}\left(\boldsymbol{R}+\lambda\boldsymbol{s}\right)}\Phi_{\boldsymbol{s}}\left(\boldsymbol{r}-\boldsymbol{R}-\boldsymbol{s}\right)$.
For periodic gauge $\overline{\lambda}=1-\lambda=1$ and for relative
gauge $\overline{\lambda}=0$, which implies that the latter yields
the simplest formulas.

In the context of link variables this will also lead to modifications.
Calculating the full expression and using the approximation for the
basis connection we get for infinitesimally close $\boldsymbol{k}$-points
(hiding $\boldsymbol{s}$ and $\mu$ sums)
\begin{align*}
\left\langle u^{\boldsymbol{k}}\mid u^{\boldsymbol{k}+\boldsymbol{h}}\right\rangle  & =\left\langle \mathrm{e}^{-i\boldsymbol{k}r}\Psi^{\boldsymbol{k}}\mid\mathrm{e}^{-i\left(\boldsymbol{k}+\boldsymbol{h}\right)\boldsymbol{r}}\Psi^{\boldsymbol{k}+\boldsymbol{h}}\right\rangle \\
 & \approx\left\langle \Psi^{\boldsymbol{k}+}\mid\mathrm{e}^{i\boldsymbol{k}\boldsymbol{r}}\left(1-ii\left(\boldsymbol{h}\boldsymbol{\nabla}_{\boldsymbol{k}}\right)\right)\mathrm{e}^{-i\boldsymbol{k}\boldsymbol{r}}\Psi^{\boldsymbol{k}}\right\rangle \\
 & =1-i\left\langle \Phi^{\boldsymbol{k}+}\mid\boldsymbol{\beta}_{\boldsymbol{k}}\Phi^{\boldsymbol{k}}\right\rangle \boldsymbol{h}\\
 & =1-iC^{\boldsymbol{k}+}\left(\overline{\lambda}S^{\boldsymbol{k}}\boldsymbol{s}+i\boldsymbol{\nabla}_{\boldsymbol{k}}\right)C^{\boldsymbol{k}}\boldsymbol{h}
\end{align*}

Dropping the basis connection is equivalent to replacing $\left\langle u^{\boldsymbol{k}}\mid u^{\boldsymbol{k}+\boldsymbol{h}}\right\rangle $
by $D^{\boldsymbol{k}+}D^{\boldsymbol{k}}$ with an appropriately
chosen $D^{\boldsymbol{k}}=\varphi^{\boldsymbol{k}}C^{\boldsymbol{k}}$
containing a phase factor $\varphi^{\boldsymbol{k}}$.

For topology a periodic wave function $\Psi^{\boldsymbol{k}+\boldsymbol{G}}=\Psi^{\boldsymbol{k}}$
is required. In detail for the Bloch sums $\Phi_{\boldsymbol{s}}^{\boldsymbol{k}+\boldsymbol{G},\lambda}=\Phi_{\boldsymbol{s}}^{\boldsymbol{k}\lambda}\mathrm{e}^{i\boldsymbol{G}\lambda\boldsymbol{s}}$
and periodicity requires $C_{\boldsymbol{s}n}^{\boldsymbol{k}+\boldsymbol{G}}=\mathrm{e}^{-i\boldsymbol{G}\lambda\boldsymbol{s}}C_{\boldsymbol{s}n}^{\boldsymbol{k}}$:
whenever a wave vector $\boldsymbol{k}+\boldsymbol{G}$ occurs in
an algorithm for which $C_{\boldsymbol{s}n}^{\boldsymbol{k}}$ is
already in use one needs to use $\mathrm{e}^{-i\boldsymbol{G}\lambda\boldsymbol{s}}C_{\boldsymbol{s}n}^{\boldsymbol{k}}$
instead of $C_{\boldsymbol{s}n}^{\boldsymbol{k}+\boldsymbol{G}}$.

From Ref.~\onlinecite{Koe23} we know that the relative gauge $\overline{\lambda}=0$
leads to vanishing approximations for $\boldsymbol{A}_{\Phi}^{\boldsymbol{k}}$
and hence we choose $\varphi_{\boldsymbol{s}}^{\boldsymbol{k}}=\mathrm{e}^{-i\boldsymbol{k}\overline{\lambda}\boldsymbol{s}}$,
since then $\Phi_{\boldsymbol{s}}^{\boldsymbol{k}\lambda}=\Phi_{\boldsymbol{s}}^{\boldsymbol{k},\overline{\lambda}=0}\varphi_{\boldsymbol{s}}^{\boldsymbol{k}}$
and we are allowed to drop $\Phi_{\boldsymbol{s}}^{\boldsymbol{k},\overline{\lambda}=0}$
(which is in relative gauge). Consequently we need to modify $C_{\boldsymbol{s}\mu,n}^{\boldsymbol{k}}\to D_{\boldsymbol{s}\mu,n}^{\boldsymbol{k}}=\mathrm{e}^{-i\boldsymbol{k}\overline{\lambda}\boldsymbol{s}}C_{\boldsymbol{s}\mu,n}^{\boldsymbol{k}}$
when calculating link variables from coefficient matrices only. For
orthogonal $\Phi^{\boldsymbol{k}}$ ($S^{k}=1$) one gets ($s,\mu$-sum implied)
\begin{align*}
D^{\boldsymbol{k}+}D^{\boldsymbol{k}} = 
D_{s\mu,n^{\prime}}^{\boldsymbol{k}+}D_{s\mu,n}^{\boldsymbol{k}+\boldsymbol{h}} & =C_{\boldsymbol{s}\mu,n^{\prime}}^{\boldsymbol{k}+}\mathrm{e}^{-i\boldsymbol{h}\overline{\lambda}\boldsymbol{s}}C_{\boldsymbol{s}\mu,n}^{\boldsymbol{k}+\boldsymbol{h}}\\
 & \approx1-iC_{\boldsymbol{s}\mu,n^{\prime}}^{\boldsymbol{k}+}\left(\overline{\lambda}\boldsymbol{s}+i\boldsymbol{\nabla}_{\boldsymbol{k}}\right)C_{\boldsymbol{s}\mu,n}^{\boldsymbol{k}}\boldsymbol{h}
\end{align*}
which is the same as derived above. In conclusion we suggest to either
use the relative gauge $\overline{\lambda}=0$ or the additional phase
factor when calculating link variables based on coefficient matrices only.

\section*{AUTHOR CONTRIBUTIONS}
G.B.A. performed DFT calculations, Wannier mapping and search for Weyl points, and wrote the first draft of the manuscript;
M.R. performed DFT calculations and revised the manuscript;
K.K. wrote the Appendix and provided the code for the calculation of the Berry phase on a closed loop.
M.P.G. planned the project, performed DFT and AHC calculations and revised the manuscript. All authors proof-read the manuscript.

\bibliography{ref}
\vspace{9cm}

\end{document}